\documentclass[nohyperref]{article}

\usepackage{microtype}
\usepackage{graphicx}
\usepackage{booktabs} 

\usepackage{caption}
\usepackage{subcaption}

\usepackage{hyperref}

\usepackage[accepted]{icml2022}

\usepackage{amsmath}
\usepackage{amssymb}
\usepackage{mathtools}
\usepackage{amsthm}

\usepackage[capitalize,noabbrev]{cleveref}

\theoremstyle{plain}

\theoremstyle{definition}

\theoremstyle{remark}

\usepackage[textsize=tiny]{todonotes}

\icmltitlerunning{Trustworthy Transformers for COVID-19 Screening With Chest Radiography}

\begin{document}

\twocolumn[
\icmltitle{Towards Trustworthy Healthcare AI: Attention-Based Feature Learning for COVID-19 Screening With Chest Radiography}

\icmlsetsymbol{equal}{*}

\begin{icmlauthorlist}
\icmlauthor{Kai Ma}{equal,uwaterloo}
\icmlauthor{Pengcheng Xi}{equal,uwaterloo,nrc}
\icmlauthor{Karim Habashy}{uwaterloo,nrc}
\icmlauthor{Ashkan Ebadi}{uwaterloo,nrc}
\icmlauthor{Stéphane Tremblay}{nrc}
\icmlauthor{Alexander Wong}{uwaterloo}
\end{icmlauthorlist}

\icmlaffiliation{uwaterloo}{Faculty of Engineering, University of Waterloo, Waterloo, Ontario, Canada}
\icmlaffiliation{nrc}{Digital Technologies Research Centre, National Research Council Canada, Ottawa, Ontario, Canada}

\icmlcorrespondingauthor{Kai Ma}{k78ma@uwaterloo.ca}
\icmlcorrespondingauthor{Pengcheng Xi}{pengcheng.xi@nrc-cnrc.gc.ca}

\icmlkeywords{Machine Learning, ICML}

\vskip 0.3in
]

\printAffiliationsAndNotice{\icmlEqualContribution} 

\begin{abstract}
Building AI models with trustworthiness is important especially in regulated areas such as healthcare. In tackling COVID-19, previous work uses convolutional neural networks as the backbone architecture, which has shown to be prone to over-caution and overconfidence in making decisions, rendering them less trustworthy --- a crucial flaw in the context of medical imaging. In this study, we propose a feature learning approach using Vision Transformers, which use an attention-based mechanism, and examine the representation learning capability of Transformers as a new backbone architecture for medical imaging. Through the task of classifying COVID-19 chest radiographs, we investigate into whether generalization capabilities benefit solely from Vision Transformers' architectural advances. Quantitative and qualitative evaluations are conducted on the trustworthiness of the models, through the use of ``trust score" computation and a visual explainability technique. We conclude that the attention-based feature learning approach is promising in building trustworthy deep learning models for healthcare.

\end{abstract}

\section{Introduction}
\label{sec:intro}

COVID-19 has affected lives around the world and continues with Omicron XE variant and possible future mutants, variants or recombinants. Screening and monitoring of patients regarding the virus have never been more important. In dealing with the pandemic, building AI models with trust is an important topic for researchers, clinicians, and patients.

The medical image analysis field has been dominated by deep Convolutional Neural Networks (CNN). A drawback of deep CNNs is that they are often overly cautious for correct predictions of the minority class, while exhibiting overconfidence for incorrect predictions of the majority class \cite{ghost}. This problem is significant in medical image analysis, considering that results with low trust may lead to serious medical consequences.

Recent innovations in computer vision (CV) have shown that Vision Transformer models can outperform CNNs. The models are based on an attention mechanism, first introduced for natural language processing (NLP) \cite{transformer}. In \cite{vit}, the authors introduced the original Vision Transformer named ViT, which yielded modest performance when trained on ImageNet \cite{imagenet_cvpr09}, with limited generalization capabilities \cite{deit}. Based on this architecture, a hierarchical variant named Swin Transformer \cite{liu2021Swin} was proposed to compute representations with shifted windows. The hierarchical architecture allows flexibility to model at various scales besides computational efficiency. The Swin Transformer has achieved state-of-the-art (SOTA) performance on tasks including image classification, object detection, and semantic segmentation. It can therefore serve as a general-purpose backbone for computer vision.

There is potential for applying Transformers to medical image analysis. CNNs have inductive biases, such as locality and translational equivariance, baked into their architectures; however, Transformers use self-attention to establish relationships between long-range elements and can attend over different regions of an image, leading to the integration of information across the entire image. When radiologists or doctors read chest X-rays, they tend to read different parts of the images and combine features for reporting and diagnosis \cite{radiography}, which is analogous to a Transformer's self-attention. Furthermore, the attention mechanism used in Transformers has an improved ability to preserve spatial information, which can lead to improved performance and trust in models \cite{transformers_see}.

In this study, we propose an attention-based model architecture to improve model trustworthiness, through leveraging the unique strength of Vision Transformers. For validations, we choose ResNet-50 \cite{resnet} and DenseNet-121 \cite{densenet} for comparison as they are commonly used as baselines and have also demonstrated good performance on similar tasks \cite{covidresnet1,covidresnet2}. The vanilla Swin Transformer is chosen as a representative Vision Transformer model, as it represents a step up from the modest original ViT but is also not so architecturally advanced as to lead to an unfair comparison. 

We study model trustworthiness with quantitative measures and qualitative evaluations. Using CNNs or Transformers as backbones, the models are pre-trained on ImageNet-1K and fine-tuned on COVIDx \cite{Wang2020}, a chest X-ray dataset which combines several well-known public data repositories such as RSNA \cite{rsna, RAHMAN2021104319} and ActualMed \cite{actualmed}. We adopt a novel trust score introduced in \cite{trust_wong} and an explainability technique named Ablation-CAM \cite{ablation-cam}. Experimental results indicate that attention-based models are consistently more trustworthy than CNN counterparts with similar performance. To our best knowledge, this is the first evaluation of Vision Transformers using the quantitative trust scores, indicating a feasibility of building trustworthy AI models for healthcare.

\section{Literature Review}
\label{sec:litrev}
Vision Transformers (ViT) were first introduced in \cite{vit}, using image patches with positional embeddings. The key mechanism behind the architecture is known as ``attention", which aims to emphasize important parts of input data while diminishing others for the purpose of imitating human cognitive attention. With the inaugural success of the ViT models, several variations have followed. A notable development is the Swin Transformer \cite{liu2021Swin},  which uses a shifted windowing scheme: by attention to local windows, it allows for cross-window connections and implements a merging layer to combine image patches. It has therefore achieved high efficiency and exceptional performance, surpassing SOTA CNNs. 

Other developments, including CoAtNet \cite{coatnet}, MaxViT \cite{maxvit}, and SwinV2 \cite{swinv2}, have consistently shown high performance on ImageNet benchmarks, beating previous approaches. These models are available in different versions, while pre-trained on different datasets. For example, the Swin Transformer has variants pre-trained on either ImageNet-1K or ImageNet-22K \cite{imagenet_cvpr09}, base version Swin-B, small versions Swin-T and Swin-S, and large version Swin-L \cite{liu2021Swin}), as well as using different image sizes.

COVID-19 causes the density of the lungs to increase, often presenting as whiteness in the lungs when seen in chest radiographs. This may obscure normal lung markings, resulting in a ``ground glass" pattern \cite{radiography}. In severe cases, lung markings can be completely lost due to whiteness, which is called ``consolidation". Furthermore, coarse, horizonal white lines can appear, which are termed ``peripheral linear opacities". When examining chest radiographs for COVID-19, radiologists review images systematically, including examining the heart, lungs, ribs, diaphragm, and mediastinum for evidence of ground glass opacity, peripheral linear opacities, or consolidation. While these are commonly used as evidence for diagnosis, changes due to COVID-19 can be subtle or even absent.

Deep learning metrics are well-documented for evaluating performance, efficiency, and explainability \cite{explainabilitymetric,robustnessmetric,netscore}. In contrast, model trustworthiness is a relatively new area of interest. Most approaches focus on quantifying trust for a prediction made on a single data sample, using uncertainty estimations \cite{uncertprop,uncertestimation} or agreement with a modified nearest-neighbour classifier \cite{trustneighbour}. However, these methods are often difficult to interpret and have high complexity \cite{uncertbayesian}.

\begin{figure*}[ht]
    \begin{center}
    \includegraphics[scale=0.39]{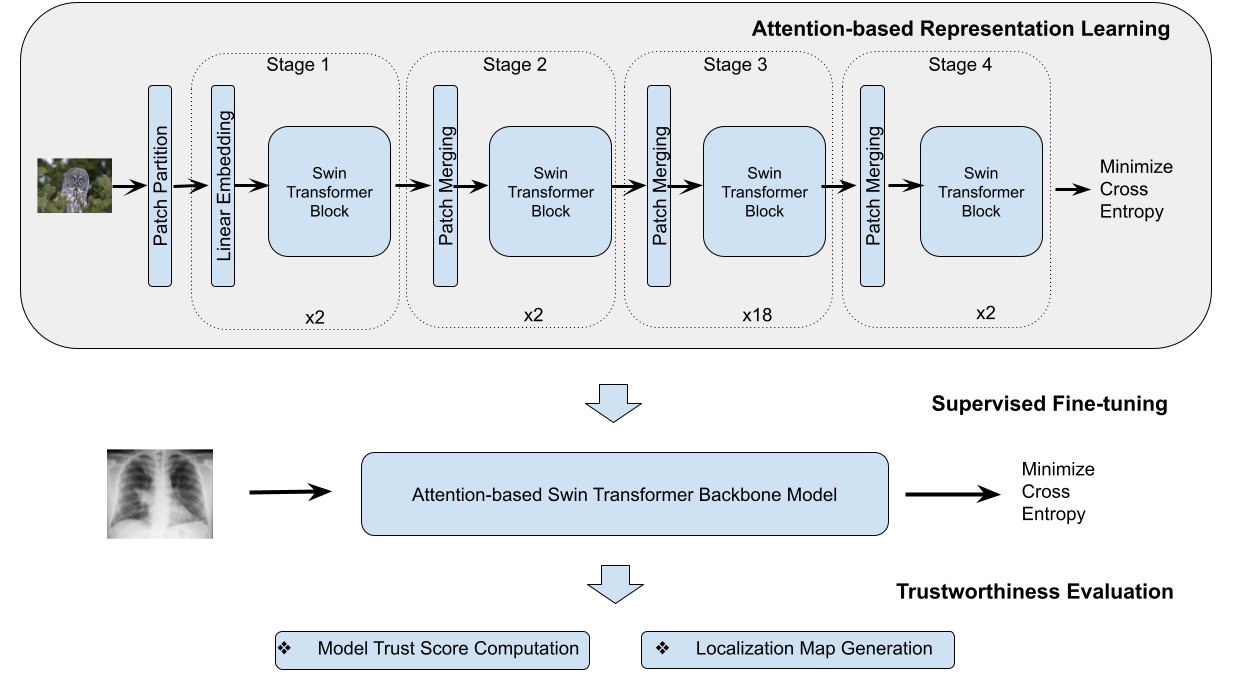}
    \caption{Model architecture and fine-tuning pipeline.}
    \label{fig:architecture}
    \end{center}
\end{figure*}

To address the above issue, authors in \cite{trust_wong} introduced the concept of ``question-answer trust", where the trustworthiness of a deep neural network is determined by its behaviour under correct and incorrect answer scenarios. To that end, the question-answer trust fundamentally penalizes undeserved confidence while rewarding well-placed confidence. A simple scalar ``trust score" metric is introduced to put question-answer trust to practice --- a higher trust score indicates a more trustworthy model. This trust score method was tested experimentally with ImageNet, achieving insightful results; furthermore, it has been adopted for medical applications \cite{Meng2022, Parra2021-lj}, showing its validity and applicability to the task at hand.

For model explainability, a common approach is to locate regions in a given input image that are used by the model for classifications. There is existing literature on this topic for Transformers in the context of medical imaging \cite{jcm11113013}; however, the authors use attention maps based on self-attention score, which is specific to Transformers and does not apply to CNNs, and no comparison is done between the two. 

We aim to use an approach that is applicable to both CNNs and Transformers. One example approach for producing localization maps is Grad-CAM \cite{grad-cam}, which uses gradients flowing into the final layer. Due to this focus on the final layer, Grad-CAM can be ineffective for Transformer architectures, which have relatively uniform representations across all layers \cite{transformers_see}. Furthermore, Grad-CAM can be prone to low-quality visualizations as a result of gradient saturation \cite{ablation-cam}. Instead, another approach named Ablation-CAM \cite{ablation-cam} solves these problems by using ablation analysis instead of gradients, resulting in immunity to saturation. As such, Ablation-CAM is able to provide visualizations that are more complete and trustworthy, and suitable for both Transformers and CNNs, demonstrating better performance than Grad-CAM for our purposes.

\section{Methodology}
\label{sec:method}
Our methodology is motivated by the unique architecture of Vision Transformers and their capability of creating models with improved trustworthiness. In particular, Swin Transformer \cite{liu2021Swin} was introduced to prove its capability of serving as a general-purpose backbone for computer vision. It is a hierarchical transformer which computes representations with shifted windows, bringing efficiency by limiting self-attention computation to non-overlapping local windows. Its key design element lies in the shift of the window partitions between consecutive self-attention layers.

\subsection{Model Architecture}\label{subsec:arch}
We propose the model architecture in Fig. \ref{fig:architecture}, where we follow a transfer learning approach \cite{bigTransfer}. We first conduct attention-based representation learning to build the backbone transformer model. We then freeze the encoder part and only fine-tune the classifier parts of the Swin Transformer models using the COVIDx data set. We specifically use the Swin-Base Transformer model, henceforth referred to as ``Swin-B". Models are created using TorchVision \cite{torchvision}, then modified such that the final fully connected layer is replaced with a single output neuron for binary classification. To produce probability distributions, a sigmoid layer is applied over the raw logits.

Validation is conducted after each training epoch and a classification threshold is calculated by maximizing validation F1-score. The model achieving the best validation accuracy is saved and then evaluated on the unseen test split. Finally, trustworthiness of the saved model is evaluated by calculating a ``trust score" for the positive class according to the method described in \cite{trust_wong}. This metric expresses model trustworthiness by rewarding well-placed confidence and penalizing undeserved overconfidence. A higher trust score indicates that a model is more trustworthy.

\subsection{Trust Score Computation} \label{subsec:trustcomp}
Trust scores are computed using the method introduced in \cite{trust_wong}. Given a question $x$, an answer $y$ given by a model $M$, such that $y = M(x)$, and $z$ representing the correct answer to $x$, we use $ R_{y=z \mid M}$ to denote the space of all questions where the answer $y$ given by model $M$ matches the correct answer $z$. Similarly, $R_{y \neq z \mid M}$ is used to denote the space of all questions where $y$ does not match the correct answer. The confidence of $M$ in an answer $y$ to question $x$ is denoted as $C(y\mid x)$. Thus, the \textit{question-answer trust} of an answer $y$ given by model $M$ for a question $x$, with knowledge of the correct answer $z$, is defined as
\[  Q_z(x,y) =  \left\{
\begin{array}{ll}
      C(y\mid x)^\alpha, & \text{if } x \in R_{y=z \mid M} \\
      (1 - C(y\mid x))^\beta, & \text{if } x \in R_{y\neq z \mid M}, \\
\end{array}
\right. \]
with $\alpha$ and $\beta$ denoting reward and penalty coefficients.

In practical usage with our models, we first calculate an optimal threshold value by maximizing F1-score on the validation split. Samples are individually passed through the model; outputs below the threshold are predicted to be negative, while outputs above the threshold are predicted to be positive. We then normalize outputs, such that negative predictions are scaled between 0 and 0.5 and positive predictions are scaled between 0.5 and 1. This allows us to represent model confidence, which is then used to compute a \textit{trust score} according to the method introduced above. To equally reward well-placed confidence and undeserved overconfidence, we use $\alpha = 
\beta = 1$. Finally, an overall positive class trust score for the model is determined by calculating the mean of all individual scores for all of the positive samples in the unseen test split.

\section{Experimental Results} \label{sec:result}

\subsection{Dataset}
\label{subsec:datasets}
For a fair comparison, all the models are pre-trained on the same dataset ImageNet-1K \cite{imagenet_cvpr09}. We fine-tune the models on COVIDx (Version 9B) \cite{Wang2020}, which consists of 30,482 chest radiographs with annotations on having COVID-19 or not (Table \ref{tab:datatab}). The dataset has a defined testing subset of 400 images, which is used for evaluating the fine-tuned models. Within the training split, 10\% of the images are sampled for validations.

\begin{table}[ht]
\caption{Data split for COVIDx V9B}\label{tab:datatab}
\begin{sc}
\begin{center}
\vskip 0.15in
\begin{tabular}{l c c c}
\toprule
\textbf{Split} & \textbf{Negative}  & \textbf{Positive} & \textbf{Total}\\
\midrule
Train    & 13,992   & 15,950  & 30,482  \\
Test     & 200      & 200     & 400  \\
\bottomrule
\end{tabular}
\end{center}
\end{sc}
\vskip -0.1in
\end{table}

\subsection{Experiment Setups}\label{subsec:setups}
Input images are resized and center cropped to 224$\times$224 and normalized, with random horizontal flipping as the only data augmentation being implemented. More data augmentations, while being effective for improving model performance \cite{deit}, will complicate the strict architecture-based comparison we are conducting.

Cross-entropy loss with SGD optimizer is used with a weight decay of $1e-4$ and momentum of $0.9$. Initial learning rates are set for the different number of training epochs (Table \ref{tab:lrset}). Cosine annealing learning rate decay is used.

\begin{table}[ht]
\caption{Initial learning rates by number of training epochs}\label{lr}%
\begin{sc}
\begin{center}
\vskip 0.15in
\begin{tabular}{ l  c  c  c  c }\hline
\toprule
\textbf{Epochs} & 30 & 50 & 100 & 200 \\
\midrule
\textbf{Initial Lr} & 5$e$-4 & 5$e$-4 & 3$e$-4 & 1$e$-4 \\
\bottomrule
\end{tabular}
\label{tab:lrset}
\end{center}
\end{sc}
    \vskip -0.1in
\end{table}

\subsection{Model Performance and Trust} \label{subsec:perftrust}
To establish a baseline result for the dataset, we fine-tune ResNet-50 and DenseNet-121 for 200 epochs and compute trust scores. Our best model fine-tuned on the Swin Transformer backbone is trained with 200 epochs. In Tables \ref{tab:precision-table}, \ref{tab:sensitivity-table}, and \ref{tab:trust-table}, we also include the precision, sensitivity, and trust scores of models trained with less epochs.

\begin{table}[ht]\label{precision}
\begin{sc}
    \begin{center}
    \caption{Precision scores on the unseen COVIDx V9B test split. The
    best results in each class are bolded.}
    \vskip 0.15in
    \begin{tabular}{@{} l  c  c @{}}
        \toprule
        \textbf{Model} & \textbf{Negative}  & \textbf{Positive} \\
        \midrule
        ResNet (200 epochs)      & \textbf{0.952}   & \textbf{1.000}   \\
        DenseNet (200 epochs)    & 0.948             & 0.995     \\
        \midrule
        Swin-B (30 epochs)      & 0.926            & \textbf{1.000}   \\
        Swin-B (50 epochs)      & 0.935            & \textbf{1.000}  \\
        Swin-B (100 epochs)     & 0.930            & \textbf{1.000}   \\
        Swin-B (200 epochs)     & \textbf{0.952}   & \textbf{1.000}   \\
        \bottomrule
    \end{tabular}
    \label{tab:precision-table}
    \end{center}
\end{sc}
    \vskip -0.1in
\end{table}

\begin{table}[ht]\label{sensitivity}
\begin{sc}
    \begin{center}
    \caption{Sensitivity scores on the unseen COVIDx V9B test split. The
    best results in each class are bolded.}
    \vskip 0.15in
    \begin{tabular}{@{} l  c  c @{}}
        \toprule
        \textbf{Model} & \textbf{Negative}  & \textbf{Positive} \\
        \midrule
        ResNet (200 epochs)      &\textbf{1.000}    & \textbf{0.950}   \\
        DenseNet (200 epochs)    & 0.995             & 0.945     \\
        \midrule
        Swin-B (30 epochs)      & \textbf{1.000}   & 0.920  \\
        Swin-B (50 epochs)      & \textbf{1.000}   & 0.930   \\
        Swin-B (100 epochs)     & \textbf{1.000}   & 0.925  \\
        Swin-B (200 epochs)     & \textbf{1.000}   & \textbf{0.950}  \\
        \bottomrule
    \end{tabular}
    \label{tab:sensitivity-table}
    \end{center}
\end{sc}
    \vskip -0.1in
\end{table}

\begin{table}[ht]\label{trust}
\begin{sc}
    \begin{center}
    \caption{Trust scores calculated from each experiment on the positive class. The
    best result is bolded.}
    \vskip 0.15in
    \begin{tabular}{@{} l  c @{}}
        \toprule
        \textbf{Model} & \textbf{Trust Score}  \\
        \midrule
        ResNet (200 epochs)      & 0.923     \\
        DenseNet (200 epochs)    & 0.922    \\
        \midrule
        Swin-B (30 epochs)      & 0.943    \\
        Swin-B (50 epochs)      & 0.959     \\
        Swin-B (100 epochs)     & 0.954    \\
        Swin-B (200 epochs)     & \textbf{0.963}    \\
        \bottomrule
    \end{tabular}
    \label{tab:trust-table}
    \end{center}
\end{sc}
\end{table}

From the results in Tables \ref{tab:precision-table} and \ref{tab:sensitivity-table}, we see that the Swin-B models are consistently matching ResNet-50 and surpassing DenseNet-121 for the precision metric of the positive class and the sensitivity metric of the negative class. These metrics are especially meaningful in the context of medical imaging: a higher sensitivity score in the negative class implies that there are less false-positives. Moreover, the Swin-B model (200-epoch) was able to match the baseline ResNet-50 model in both classes for precision and sensitivity. This highly successful fine-tuned Swin-B model will be made publicly available upon acceptance of this work.

While the ResNet-50 model outperforms some Swin-B models (trained for less than 200 epochs) in terms of precision and recall, the results presented in table \ref{tab:trust-table} indicate that Swin-B models consistently achieve significantly higher trust scores than both ResNet-50 and DenseNet-121. The aforementioned Swin-B model (200-epoch) achieved the highest trust score out of all models. Thus, it is shown that Swin Transformers can achieve competitive performance to ResNet-50 and surpass DensNet-121 while being significantly more trustworthy.

We conducted additional experiments to study representation learning capabilities of vanilla Swin Transformers. Specifically, fine-tuning was performed on a smaller variant of the Swin Transformer, the Swin-Tiny (Swin-T), and it is compared with the Swin-B model when both are fine-tuned for 50 epochs. Results are listed in Tables \ref{tab:add-precision-table}, \ref{tab:add-sensitivity-table}, and \ref{tab:add-trust-table}. Although the Swin-T model is lackluster in terms of model performance, it achieved a trust score similar to Swin-B model, which is much higher than the ResNet-50 and DenseNet-121 baselines. This strengthens our conclusion that the Transformer architecture itself can contribute to model trustworthiness. 

\begin{figure}[ht] \label{figure:explainability}
    \centering
    \begin{subfigure}{1\linewidth}
        \centering
        \includegraphics[scale=0.33]{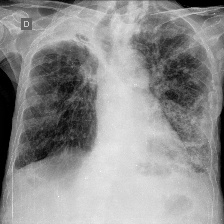}
        \includegraphics[scale=0.33]{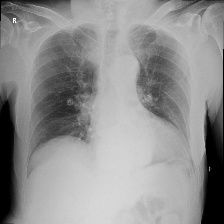}
        \includegraphics[scale=0.33]{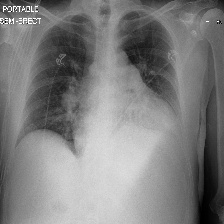}
        \caption{Original chest radiographs for positive COVID-19 samples \break}
        \label{fig:original-cam}
    \end{subfigure}
    \begin{subfigure}{1\linewidth}
        \centering
        \includegraphics[scale=0.33]{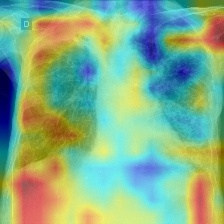}
        \includegraphics[scale=0.33]{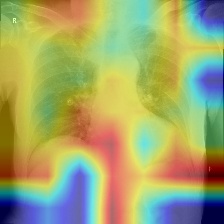}
        \includegraphics[scale=0.33]{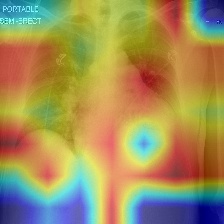}
        \caption{Swin-B 200-epoch Ablation-CAM \break}
        \label{fig:swin-cam}
    \end{subfigure}
    \begin{subfigure}{1\linewidth}
        \centering
        \includegraphics[scale=0.33]{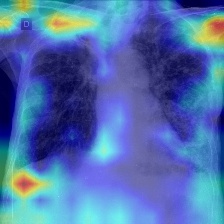}
        \includegraphics[scale=0.33]{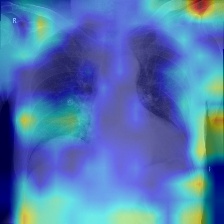}
        \includegraphics[scale=0.33]{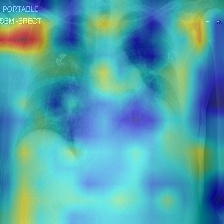}
        \caption{ResNet-50 200-epoch Ablation-CAM}
        \label{fig:resnet-cam}
    \end{subfigure}
    \caption{Swin-B and ResNet-50 Ablation-CAMs for 3 selected COVID-positive chest X-rays. ResNet-50 is chosen as a representative because it produced better results and localization maps than Densenet-121. Warm colors (red, orange) indicate high importance, cold colors (blue, green) indicate lower importance.}
\end{figure}

\begin{table}[ht] 
\begin{sc}
    \begin{center}
    \caption{Precision scores on the unseen COVIDx V9B test split.}
    \vskip 0.15in
    \begin{tabular}{@{} l  c  c  c @{}}
        \toprule
        \textbf{Model} & \textbf{Negative}  & \textbf{Positive} \\
        \midrule
        Swin-T  & 0.913    & 1.000  \\
        Swin-B  & 0.935    &  1.000   \\
        \bottomrule
    \end{tabular}
    \label{tab:add-precision-table}
    \end{center}
\end{sc}
\vskip -0.1in
\end{table}

\begin{table}[ht]
\begin{sc}
    \begin{center}
    \caption{Sensitivity scores on the unseen COVIDx V9B test split.}
    \vskip 0.15in
    \begin{tabular}{@{} l  c  c @{}}
        \toprule
        \textbf{Model} & \textbf{Negative}  & \textbf{Positive} \\
        \midrule
        Swin-T  & 1.000 &  0.905  \\
        Swin-B & 1.000     &    0.930 \\
        \bottomrule
    \end{tabular}
    \label{tab:add-sensitivity-table}
    \end{center}
    \end{sc}
    \vskip -0.1in
\end{table}

\begin{table}[ht]
\begin{sc}
    \begin{center}
    \caption{Trust scores calculated on the positive class.}
    \vskip 0.15in
    \begin{tabular}{@{} l  c @{}}
        \toprule
        \textbf{Model} & \textbf{Trust Score}  \\
        \midrule
        Swin-T      &  0.954     \\
        Swin-B     &  0.959   \\
        \bottomrule
    \end{tabular}
    \label{tab:add-trust-table}
    \end{center}
    \end{sc}
\end{table}

\subsection{Localization Map Analysis} \label{subsec:ablationanalysis}

We produce localization maps by computing Ablation-CAMs for our models, applied to over 100 images from the COVIDx test split. These maps highlight image regions deemed important by the model for classifications. Figures \ref{fig:original-cam} and \ref{fig:swin-cam} show three example COVID-positive images and localization maps computed on the Swin-B model (200-epoch). Figure \ref{fig:resnet-cam} shows corresponding maps produced by the ResNet-50 baseline model on the same images.

The localization maps produced by Swin Transformers highlight the lung area, successfully focusing on the ground glass pattern, consolidation and peripheral linear opacities used by human doctors for diagnosis \cite{radiography}. In contrast, maps produced by the ResNet-50 are less successful, highlighting irrelevant areas such as shoulders and arms, sometimes missing the lungs entirely.

\section{Conclusion}
\label{sec:discussions}
In the context of building healthcare AI models, we propose an attention-based model architecture to improve model trustworthiness, an important topic for researchers, clinicians, and patients. We have compared the effectiveness of Transformer and CNN models used as backbone architectures for screening COVID-19 patients using chest X-ray images. Through quantitative and qualitative evaluations, we demonstrate that the attention-based models are capable of matching CNN models in performance, while achieving significantly higher trustworthiness. This advantage is further validated through localization maps, which highlight regions deemed important by the models. The Transformers are shown to be better at identifying medically relevant lung regions than CNNs. We conclude that the trustworthiness and performance obtained by the Transformers are due to their architecture improvements over the CNN counterparts; therefore, models built upon attention-based backbone architectures are promising for medical imaging tasks, including COVID-19 and other healthcare applications.

For further improvement, our future work includes fine-tuning with the larger ``Swin-L" Transformer model. Moreover, experiments with different loss functions instead of standard cross-entropy loss, such as those designed for medical image analysis \cite{auc}, will be further studied to improve performance and trust. To test the generalization capability of the proposed approach, we plan to validate the pipeline on other datasets including SIIM-FISABIO-RSNA \cite{lakhani_mongan}. Finally, while Transformer-based models are currently popular in the field of computer vision, innovations have not stopped for CNN models. We are interested in comparing Transformers to newer CNN models, such as ConvNeXt \cite{convnext}, which has achieved excellent results by incorporating Transformer-like characteristics into CNN architectures.

\bibliography{example_paper}
\bibliographystyle{icml2022}

\end{document}